\documentclass[prx,aps,twocolumn,superscriptaddress]{revtex4-2}

\usepackage[pdftex]{graphicx}
\usepackage{amsbsy,amssymb,amsmath,bm,mathtools}

\usepackage{xspace}
\usepackage{bm}
\usepackage{color}
 
\usepackage[mathscr]{euscript}

\usepackage{url}
\usepackage[section]{placeins}
\usepackage[colorlinks=true,linkcolor=blue,citecolor=blue]{hyperref}

\begin{document}



\title{Enhanced coarsening of charge density waves induced by electron correlation: Machine-learning enabled large-scale dynamical simulations}

\author{Yang Yang}
\affiliation{Department of Physics, University of Virginia, Charlottesville, VA 22904, USA}

\author{Chen Cheng}
\affiliation{Department of Physics, University of Virginia, Charlottesville, VA 22904, USA}

\author{Yunhao Fan}
\affiliation{Department of Physics, University of Virginia, Charlottesville, VA 22904, USA}

\author{Gia-Wei Chern}
\affiliation{Department of Physics, University of Virginia, Charlottesville, VA 22904, USA}

\date{\today}

\begin{abstract}
The phase ordering kinetics of emergent orders in correlated electron systems is a fundamental topic in non-equilibrium physics, yet it remains largely unexplored. The intricate interplay between quasiparticles and  emergent order-parameter fields could lead to unusual coarsening dynamics that is beyond the standard theories. However, accurate treatment of both quasiparticles and collective degrees of freedom is a multi-scale challenge in dynamical simulations of correlated electrons. Here we leverage modern machine learning (ML) methods to achieve a linear-scaling algorithm for simulating the coarsening of charge density waves (CDWs), one of the fundamental symmetry breaking phases in functional electron materials. We demonstrate our approach on the square-lattice Hubbard-Holstein model and uncover an intriguing enhancement of CDW coarsening which is related to the screening of on-site potential by electron-electron interactions. Our study provides fresh insights into the role of electron correlations in non-equilibrium dynamics and underscores the promise of ML force-field approaches for advancing multi-scale dynamical modeling of correlated electron systems.
\end{abstract}

\maketitle
\section{Introduction}

Phase ordering dynamics -- the study of how systems evolve from a disordered to an ordered state after a sudden change in external conditions -- has profound implications across fundamental physics, materials science, and technology~\cite{Bray1994,Onuki2002,Puri2009,Cugliandolo15}. Of particular interest is the dynamical evolution of emergent order parameter fields in symmetry-breaking phases. Following the pioneering works of Lifshitz, Slyozov, and Wagner~\cite{Lifshitz61,Lifshitz62,Wagner61}, great advances have been made both experimentally and theoretically in the past few decades. A unifying framework based on topological defects of order parameters has led to the classification of several universality classes of domain-growth dynamics~\cite{Bray1994}. Recent experiments have shown a wide variety of complex orders in functional electron materials, ranging from itinerant magnetism to superconductivity of various symmetries and different kinds of density waves \cite{Wu2011,Hirata2021,Park2021,Hu2022a}. Whether the standard universality classes can be applied to the coarsening dynamics of these emergent orders remains a subject of ongoing research.

Standard theories of phase ordering kinetics, such as the time-dependent Ginzburg-Landau (TDGL) equation and phase-field models, are mostly empirical field-theoretical approaches constrained only by symmetry properties and conservation laws of order parameters~\cite{Hohenberg77}. Details of the underlying electronic structures are often neglected in these general approaches. However, the intricate interplay between quasiparticles and collective order-parameter degrees of freedom in correlated electron systems could give rise to unusual coarsening behaviors and non-equilibrium dynamics, in general. For example, it has been argued that coupling of the ferromagnetic order parameter to the fermionic soft modes leads to qualitatively new effects for the late-stage coarsening~\cite{Belitz07}. Complex domain morphologies have also been shown to result from electron-mediated long-range interactions between local-order parameters, leading to unusual domain growth phenomena~\cite{Cheng2023,Ghosh24,Fan24}. 

An accurate treatment of the electron degrees of freedom is thus crucial to capture these nontrivial effects of electrons on coarsening dynamics. This also underscores the multiscale nature of dynamical modeling for functional electron materials. The simulations must account for distinct time scales, as the collective order-parameter fields typically evolve on a slower time scale compared to the underlying quasiparticles. Furthermore, large-system simulations are necessary in order to minimize finite-size effects on pattern formation and domain coarsening during phase-ordering processes. This also means that an inhomogeneous electronic structure problem has to be solved at each time step in order to obtain the electronic driving forces acting on the order-parameter field. Moreover, time-consuming many-body calculations are often required in order to properly account for electron correlation effects. As the computational complexity of most many-body methods is super-linear, conventional approaches to large-scale dynamical simulation are unattainable.

Yet, as argued by W. Kohn, linear-scaling electronic structure methods are possible provided the so-called ``nearsightedness principle" is satisfied~\cite{Kohn1996,Prodan2005}. This locality principle, which is a result of wave-mechanical destructive interference, applies broadly to both insulators and metals. It relies on the presence of many particles, which do not necessarily need to interact with one another. For example, linear scalability in kernel polynomial method (KPM), a powerful technique for approximating spectral properties of large matrices, relies on computations of sparse Hamiltonian matrices which are a consequence of the locality property~\cite{Silver1994,Weisse2006,Wang2018}. However, KPM can only be used for systems without electron-electron interactions, as described by bilinear fermionic Hamiltonians. For most electronic structure methods for correlated electrons, it is unclear how to integrate the locality principle to improve the computational efficiency.

In this paper, we leverage machine learning (ML) methods to overcome the computational difficulty of large-scale dynamical simulations in correlated electron systems. In particular, we show that the locality principle can be naturally and explicitly incorporated into the ML force-field framework~\cite{behler07,bartok10,li15,shapeev16,botu17,smith17,zhang18,behler16,deringer19,mcgibbon17,suwa19,chmiela17,chmiela18,sauceda20} to achieve linear scalability. In the context of phase-ordering dynamics, the locality principle indicates that local forces that drive domain growth only depend on order-parameter configurations in the corresponding immediate surroundings. Thanks to the universal approximation theorem~\cite{Cybenko1989,Hornik1989,Barron1993}, a deep-learning neural network can then be trained to accurately capture this complex dependence on the neighborhood configurations. As the ML calculation of local forces is of a constant complexity, linear scalability of dynamical simulations can be achieved through a divide-and-conquer approach.

We demonstrate our ML approach on the phase ordering kinetics of charge density waves (CDWs), one of the prominent emergent orders in functional electron materials~\cite{Gruner1988,Gruner1994}. Although CDW systems have been intensively studied for more than two decades, the majority of works focused on their equilibrium properties and thermodynamic behaviors. The research on non-equilibrium dynamics of CDW orders is still in its infant stages, although significant experimental progress has been made in recent years thanks to the advent of ultrafast technologies~\cite{Shi2019,Maklar2021,Duan2023,Boschini2024}.  In particular, the coarsening dynamics of CDW has yet to be systematically studied, and the impact of electron correlation on the growth of CDW domains remains an open question.

As a minimal model for studying both CDW ordering and electron correlation effects, we consider the square-lattice Hubbard-Holstein (HH) model~\cite{Berger1995,Bauer10a,Bauer10b,Fehske04,Kumar2008,Weber2018}. At half electron filling, the Holstein electron-phonon coupling stabilizes a checkerboard CDW order in the adiabatic regime~\cite{Holstein1959,noack91,zhang19,chen19,hohenadler19}. On the other hand, strong on-site Hubbard repulsion~\cite{Hubbard1963} leads to localization of electrons and the emergence of N\'eel order. Although both CDW and N\'eel states are electronic insulators, their very different nature indicates the incompatibility of the two orders. This competition is also evident from the tendency of the Holstein coupling to maximize the on-site electron number, thereby enhancing the probability of double occupancy, which is energetically unfavorable to the Hubbard repulsion. 

In the context of CDW coarsening, the competition of the two mechanisms might suggest that the inclusion of a moderate Hubbard repulsion would suppress the growth of CDW domains. However, the intricate interplay of these two couplings could lead to unusual phases or dynamical phenomena. For example, convincing numerical evidences have shown that an intriguing correlated metal or potentially superconducting phase emerges in the intervening regime when the two interactions are of similar strength~\cite{Clay2005,Nowadnick12,Johnston2013,Nowadnick15,Costa2020}. Here, our extensive simulations uncover a surprising enhancement of CDW coarsening induced by electron correlation. We further show that this enhanced coarsening is related to the disorder screening phenomenon caused by electron-electron interactions.

\section{Results}

{\bf Adiabatic dynamics of CDW order} --     
The Hamiltonian for the Hubbard-Holstein (HH) model is given by
    \begin{align}\label{eqn:HH}
        \mathcal{H} =&- \sum_{ ij,\sigma} t_{ij} c^\dagger_{i,\sigma} c^{\,}_{j,\sigma}+U\sum_{i} n_{i,\downarrow} n_{i,\uparrow}\nonumber\\&+\sum_{i}\left(\frac{P_i^2}{2m}+\frac{m\omega_0^2 Q_i^2}{2}\right)-g\sum_{i,\sigma} Q_i n_{i,\sigma},
    \end{align}
where $c_{i,\sigma}^\dagger$ ($c_{i,\sigma}$) represents the creation (annihilation) operator of an electron at site $i$ with the spin $\sigma$ ($\uparrow$ or $\downarrow$), and $n_{i,\sigma}\equiv c_{i,\sigma}^\dagger c_{i,\sigma}$ represents the electron occupation number at site $i$. The lattice degree of freedom on site $i$ is described by the coordinate $Q_i$ and its conjugate momentum $P_i$. This Hamiltonian consists of three components: (1) the Hubbard model for electrons, which includes electron hopping with amplitude $t_{ij}$ and on-site Coulomb repulsion~$U$; (2) Einstein phonon model for lattice dynamics, characterized by mass~$m$ and intrinsic frequency~$\omega_0$; and (3) the Holstein coupling between the lattice and electrons, with strength~$g$. In the following we consider a regime with an adiabatic parameter $\hbar \omega_0 / t_{\rm nn} =1$ and a strong electron-phonon coupling $\lambda = g^2/K_0 t_{\rm nn} = 1.5$, where $t_{\rm nn}$ is the nearest-neighbor hopping amplitude and~$K_0 = m\omega_0^2$ is the effective elastic constant. 

\begin{figure*}[t]
\includegraphics[width=0.99\linewidth]{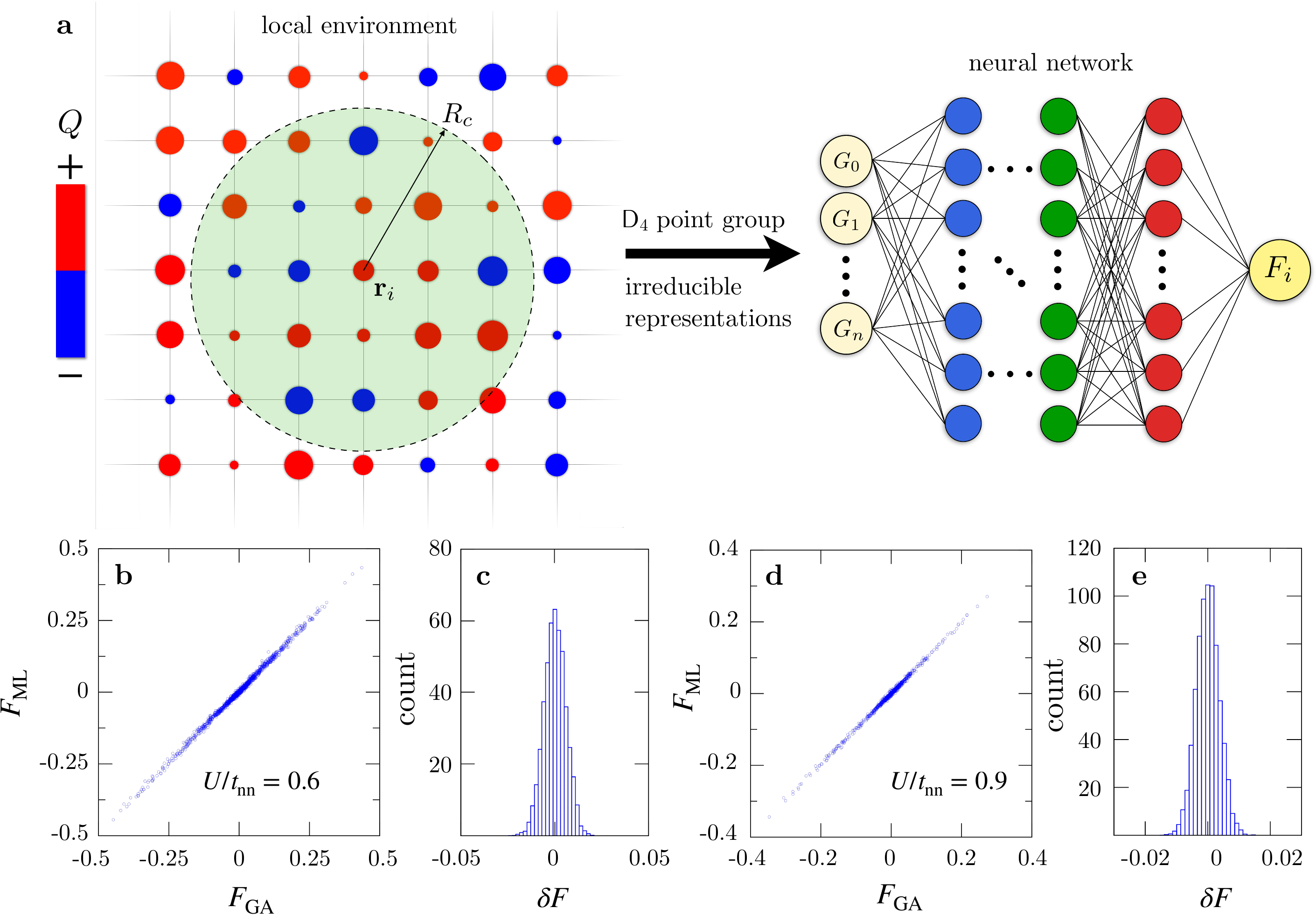}
\caption{{\bf Machine learning (ML) force-field model for the adiabatic Hubbard-Holstein model.}\\ 
{\bf a} Schematic diagram of the ML model for computing the force acting on the lattice degree of freedom $Q$ at site $\mathbf{r}_i$. The local lattice configuration $\mathcal{C}_i=\left\{Q_0,Q_1,\dots,Q_n\right\}$ is selected by the cutoff radius $R_c$. The size of disk represents the relative intensity of $Q$. The selected local lattice configuration is mapped into the symmetry-invariant descriptors $\left\{G_0,G_1,\dots,G_n\right\}$ as the input to the neural network using the irreducible representation of the point group $D_4$. The output of the neural network produces the force $F_i$. {\bf b}-{\bf e}: Benchmark for the forces computed from the ML force-field model for adiabatic dynamics of Hubbard-Holstein model. Panels {\bf b} and {\bf d} plot the ML predicted force $F_{\mathrm{ML}}$ vs the forces computed from the Gutzwiller approximation $F_{\mathrm{GA}}$ for Hubbard $U/t_{\mathrm{nn}}=0.6$ and $U/t_{\mathrm{nn}}=0.9$, respectively. The corresponding histograms, {\bf c} and {\bf e}, show the spread of the error $\delta F=F_{\mathrm{ML}}-F_{\mathrm{GA}}$. The standard deviations of the errors are $\sigma = 0.00577$ and 0.00370, respectively.}
\label{fig:1} 
\end{figure*}

Here we are interested in the strong electron-phonon coupling regime where the low-temperature phase of the HH model at half-filling exhibits a CDW order characterized by a checkerboard modulation of the electron density $n_{A/B} = 1 \pm \delta$, where $A$ and $B$ refer to the two sublattices of the bipartite lattice and $\delta$ quantifies the charge modulation. The charge modulation is accompanied by a staggered lattice distortion $Q_{A/B} = \pm \mathcal{Q}$. As a result, the CDW order breaks the $Z_2$ sublattice symmetry, representing a special type of commensurate translational symmetry breaking. From a symmetry perspective, the CDW transition is expected to belong to the Ising universality class, a prediction that is confirmed by quantum Monte Carlo (QMC) simulations~\cite{noack91,zhang19}. 

Although the Holstein model itself is amenable to QMC methods, a full quantum treatment with the inclusion of the Hubbard term is possible only for certain electron filling fractions. The HH model can also be solved within the framework of dynamical mean-field theory (DMFT) even away from half-filling. Yet both computational methods are only applicable for equilibrium properties. To make progress toward modeling the CDW coarsening,which is an intrinsically non-equilibrium process, the dynamical simulations of HH model is carried out within the semiclassical approximation, i.e. treating phonons as classical variables. This is justified, at least for CDW physics, by the fact that CDW phases obtained from hybrid Monte Carlo method based on exact diagonalization of electron Hamiltonian agree very well with the determinant QMC results~\cite{esterlis19}. Within the semiclassical approximation, the lattice degrees of freedom of the HH model is governed by the Langevin dynamics
\begin{align}\label{eqn:Langevin}
        m\frac{d^2 Q_i}{d t^2}=-\frac{\partial \langle\mathcal{H} \rangle}{\partial Q_i}-\gamma\frac{d Q_i}{d t}+\eta_i.
\end{align}
Here the Langevin thermostat is used to account for the effects of a thermal reservoir at temperature $T$ during the phase ordering; $\gamma$ is a damping constant and $\eta_i(t)$ is a thermal noise of zero mean described by correlation function $\langle \eta_i(t) \eta_j(t') \rangle = 2 m \gamma k_B T \delta_{ij} \delta(t - t')$.

The calculation of the driving force term in Eq.~(\ref{eqn:Langevin}) requires solutions of the electron density operator $\varrho(t)$ for computing the expectation value $\langle \mathcal{H} \rangle = {\rm Tr}(\varrho \mathcal{H} )$. Since domain growth during phase ordering is typically a slow process compared to the rapid relaxation of electrons, the evolution of the CDW state can be effectively described by using the adiabatic approximation, similar to the Born-Oppenheimer approximation commonly used in {\em ab initio} molecular dynamics simulations~\cite{Marx2009}. The adiabatic approximation is further justified by the fact that, in real materials with CDWs, the time scale of lattice dynamics is typically one or two orders of magnitude slower than that of electron dynamics~\cite{Hellmann2012,Sayers2020,Hu2022b}.

\begin{figure*}[t]
\includegraphics[width=0.9\linewidth]{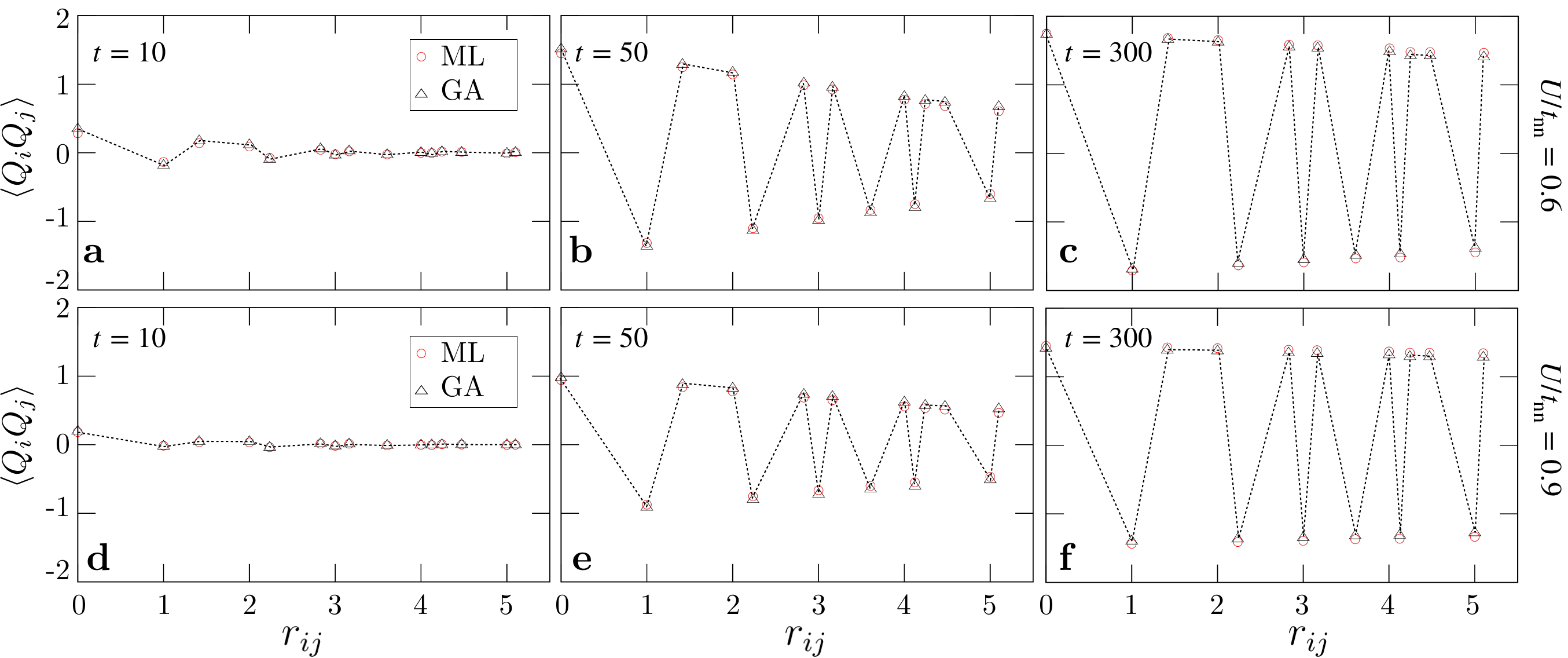}
\caption{Comparison of lattice correlation function $\langle Q_i Q_j\rangle$ obtained from Langevin simulations with the ML force-field model and the Gutswiller approximation (GA). Thermal quench simulations of a $14\times 14$ system at two different Hubbard $U/t_{\mathrm{nn}}=0.6$ ({\bf a}-{\bf c}) and $U/t_{\mathrm{nn}}=0.9$ ({\bf d}-{\bf f}) were carried out to produce these correlation functions at various time $t$ after the quench.}
\label{fig:2} 
\end{figure*}
    
\medskip    

\textbf{Machine-learning force-field approach}  -- Even with these simplifications, large-scale dynamical simulation of the HH model is still a tall order. This is because the force calculation, which is required at each time step of the dynamical simulation, requires solving a real-space Hubbard model with random on-site potentials $v_i = - g Q_i$ determined by the instantaneous phonon configuration. Depending on the many-body techniques used to solve the Hubbard model, the computational complexity of dynamical simulations ranges from polynomial $\mathcal{O}(N^c)$, with an exponent $c \ge 1$, to exponential scaling in the case of exact solution. 

As discussed above, linear-scaling algorithm can be obtained through the ML force-field approach, which was originally developed in quantum chemistry to enable large-scale {\em ab initio} molecular dynamics simulations~\cite{behler07,bartok10,li15,shapeev16,botu17,smith17,zhang18,behler16,deringer19,mcgibbon17,suwa19,chmiela17,chmiela18,sauceda20}. Similar scalable ML models have also been developed for multi-scale dynamical modeling of itinerant electron magnets and other lattice electron systems, including the Holstein model~\cite{zhang22,zhang22b,cheng23a,zhang20,zhang21,zhang23,cheng23b}. 
In the context of adiabatic dynamics of the CDW state, the computational bottleneck is the calculation of the local force $F_i \equiv -\partial \langle  \mathcal{H} \rangle / \partial Q_i$ at each lattice site. The locality principle indicates that $F_i$ only depends on structural configurations in the immediate neighborhood. This suggests a scalable ML architecture, outlined in Fig.~\ref{fig:1} {\bf a}, for the efficient prediction of the local forces. First, the lattice variables in a finite neighborhood, defined as $\mathcal{C}_i=\left\{ Q_j | |\mathbf{r}_j-\mathbf{r}_i|\leq R_c\right\}$, is selected to generate the input to the ML model. Here the cutoff radius $R_c$ is determined by the locality of the force. 

Next, in order to incorporate the discrete lattice symmetry into the ML model, the local configuration~$\mathcal{C}_i$ is transformed into a set of feature variables $\{G_1, G_2, \cdots \}$, also known as descriptors, which are invariant under operations of the $D_4$ group, the site-symmetry of the square lattice. Since two symmetry-related configurations $\mathcal{C}^{(1)}_i$ and $\mathcal{C}^{(2)}_i$ are mapped to the same descriptor, which is then fed into the neural network, it is guaranteed to produce exactly the same local force. 
This construction implies the following dependence of local force on the neighborhood configuration
\begin{eqnarray}
	F_i = \mathcal{F}\bigl( \left\{ G_1(\mathcal{C}_i), G_2(\mathcal{C}_i), \cdots \right\}\bigr).
\end{eqnarray}
Here $\mathcal{F}(\cdot)$ is a universal function which depends on the electronic structure method used to solve the HH model. Importantly, the complex dependence on the neighborhood (through the feature variables $G_l$) is to be approximated by a deep-learning neural network, which can be trained from many-body solutions on small systems. Since the size of the neighborhood is fixed, the time complexity of force prediction is of order $\mathcal{O}(1)$, independent of the system size. Linear scalability of force calculations of the whole system is achieved by repeatedly applying the same fixed-size ML model to every site of the lattice. It is important to note that the ML model outlined above is essentially a classical force-field model, yet with the accuracy of the desired quantum calculations.

\begin{figure*}[t]
\includegraphics[width=0.85\linewidth]{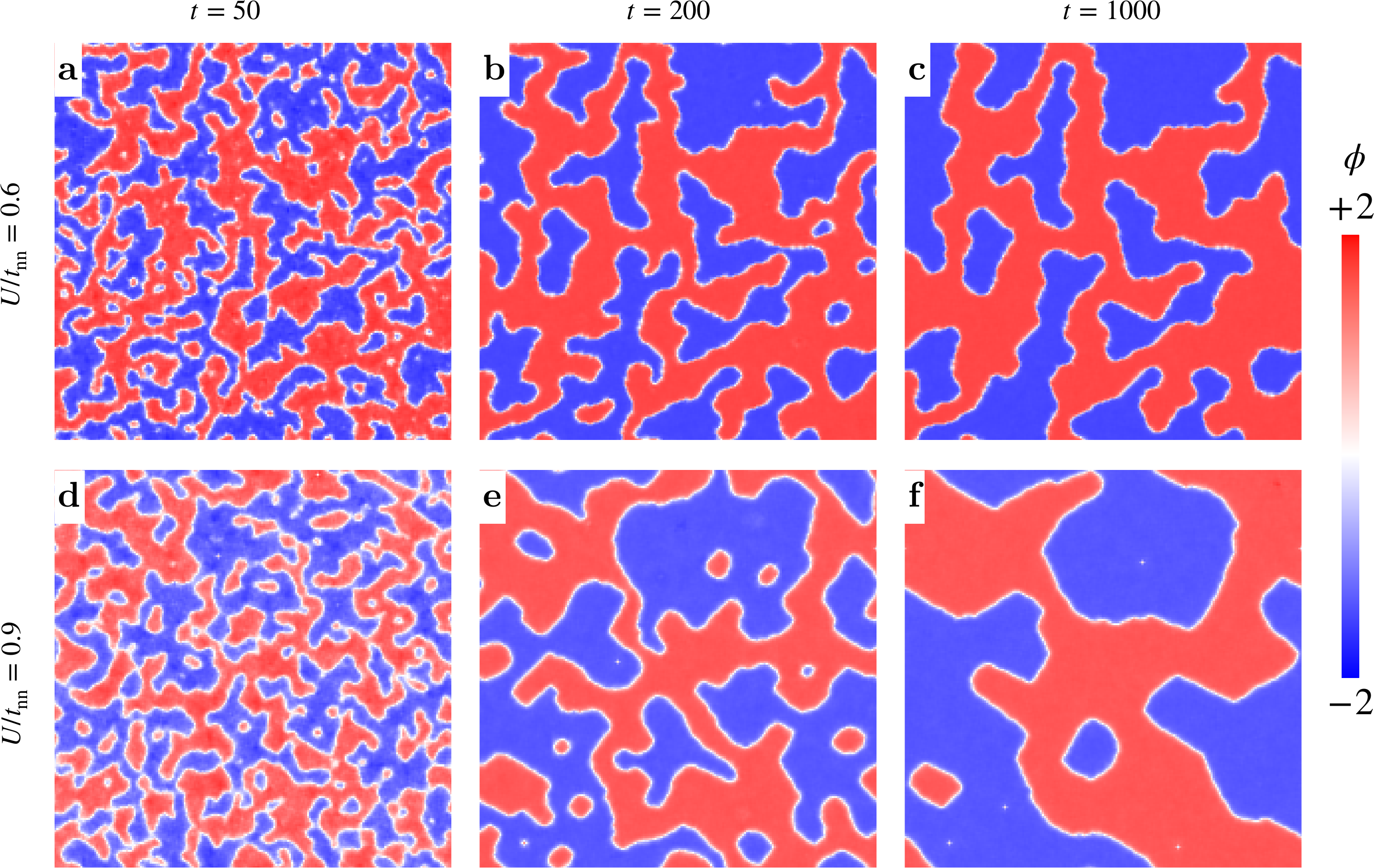}
\caption{Snapshots of local CDW parameter $\phi_i$ at various time after a thermal quench of the Holstein model for Hubbard $U/t_{\mathrm{nn}}=0.6$ ({\bf a}-{\bf c}) and $U/t_{\mathrm{nn}}=0.9$ ({\bf d}-{\bf f}). An initial random configuration is suddenly quenched to a temperature $k_B T/t_{\mathrm{nn}}= 0.001$ at time $t = 0$. The ML adiabatic Langevin dynamics is used to simulate the relaxation of the system toward equilibrium. The red and blue regions correspond to two types of CDW domains related by the $\mathbb{Z}_2$ symmetry, respectively.}
\label{fig:3} 
\end{figure*}

We note that the above ML framework can be applied to learn any given many-body calculations of a disordered Hubbard model. The training of the NN, however, still requires a large dataset from solutions on a system whose size is greater than the locality $R_c$. For more sophisticated methods, such as DQMC, the generation of dataset itself would be a time-consuming process. As a proof of principle, here we employ the Gutzwiller/slave-boson method which has proven an efficient real-space approach to disordered Hubbard models. The Gutzwiller approximation (GA) also accurately captures correlation effects such as bandwidth renormalization and disorder screening \cite{Gutzwiller1963,Gutzwiller1964,Gutzwiller1965,Bunemann1997,Bunemann1998,Bunemann2003,Fabrizio2007,Lanata2008,Lanata2012}. Moreover, thanks to its relatively high efficiency, GA can be feasibly combined with Langevin dynamics simulations on small systems. This allows us to carry out dynamical benchmarks of the trained ML model to be discussed below. 

We first benchmark the force predictions of the ML model. In Fig.~\ref{fig:1} {\bf b}-{\bf e}, we compare the forces predicted by our ML model ($F_{\mathrm{ML}}$) with those computed using the GA ($F_{\mathrm{GA}}$). The results show excellent agreement between the two methods for both Hubbard $U$ values. We also compute the standard deviation of the error $\delta F \equiv F_{\mathrm{ML}} - F_{\mathrm{GA}}$, obtaining values of $\sigma = 0.00577$ and $\sigma = 0.00370$ for $U/t_{\mathrm{nn}}=0.6$ and $U/t_{\mathrm{nn}}=0.9$, respectively.

To further validate our ML model, we compare Langevin simulations using ML predicted forces against those based on GA calculations. We initialize $Q_i$ in a random state on a $14\times 14$ square lattice at half filling, and start simulations at temperature $k_BT/t_{\mathrm{nn}}=0.001$, using two different Hubbard $U$ values: $U/t_{\mathrm{nn}}=0.6$ and $U/t_{\mathrm{nn}}=0.9$.  We compare the equal-time correlation of lattice degrees of freedom, $\langle Q_i Q_j\rangle\equiv\langle Q_i(t) Q_j(t)\rangle$, obtained from simulations using both methods. The results from Fig.~\ref{fig:2} demonstrate that the ML model accurately reproduces the same correlation as the GA method across all time windows. Notably, the equal-time correlation $\langle Q_i Q_j \rangle$ rapidly develops a staggered pattern with respect to $r_{ij}$, indicating the emergence of checkerboard CDW domains. These benchmark results give us confidence in our ML force-field model's ability to accurately simulate the coarsening dynamics of CDWs for various Hubbard $U$ values in large systems.

    \begin{figure*}[t]
	   \includegraphics[width=1.0\linewidth]{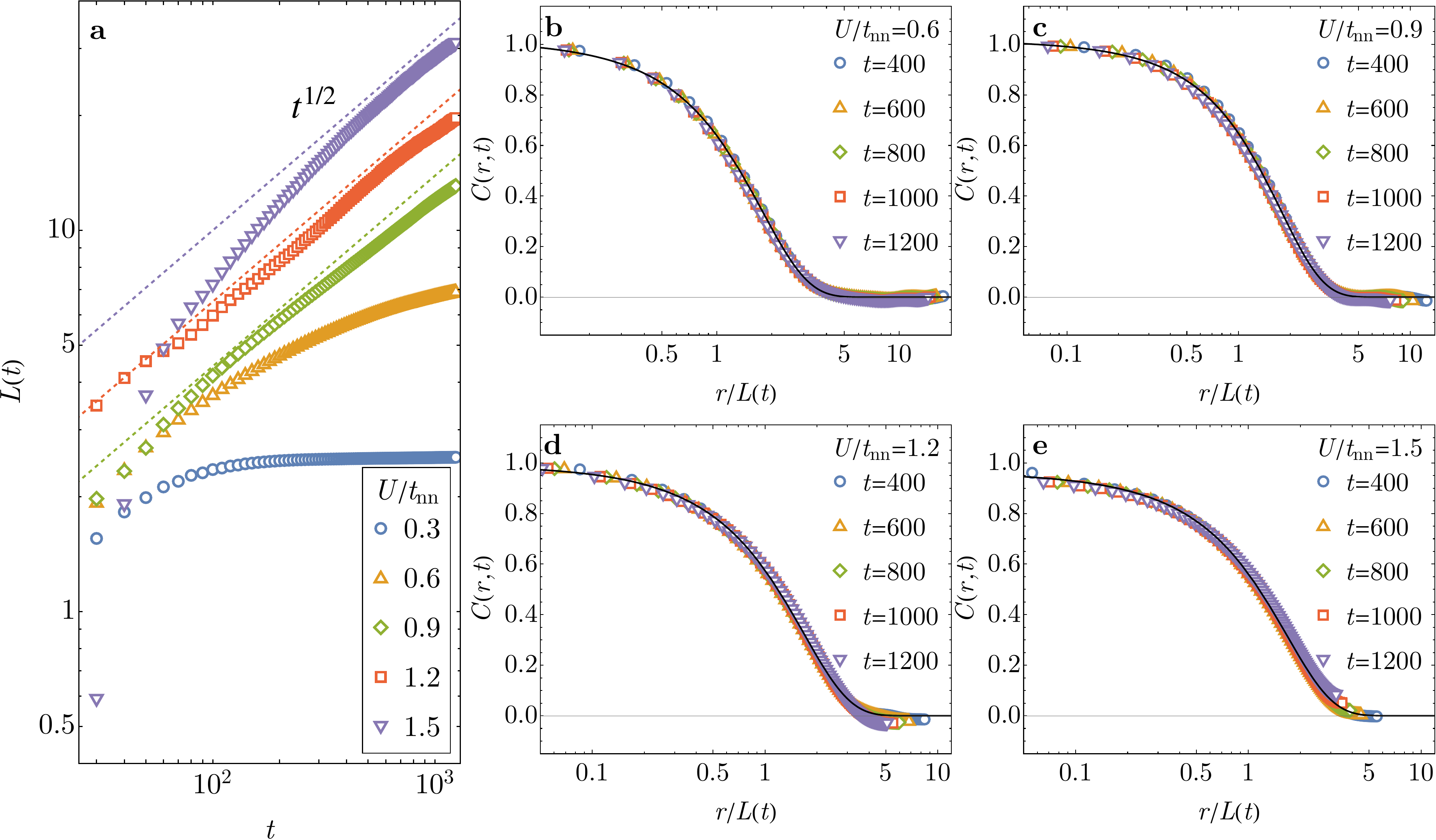}
	   \caption{{\bf a} The evolution of the characteristic length $L(t)$ of the CDW domain with respect to time for five Hubbard $U$ values ranging from $U/t_{\mathrm{nn}}=0.3$ to $U/t_{\mathrm{nn}}=1.5$. The upper three dashed lines show the growth of CDW domain recovers the Allen-Cahn growth law $L(t)\sim t^{1/2}$ (dashed line) when $U/t_{\mathrm{nn}}\geq 0.9$. 
       {\bf b}-{\bf e} Collapse of the correlation functions onto to single curve after rescaling $r/L(t)$ for $U/t_{\mathrm{nn}}=0.6$, $0.9$, $1.2$ and $1.5$.}
        \label{fig:4} 
    \end{figure*}
    
    \textbf{Coarsening dynamics of CDWs} -- To study the coarsening dynamics of CDW domains, we enlarge the system size to $200 \times 200$. We then apply our pre-trained ML force-field model in the adiabatic Langevin dynamics to compute the forces. In order to characterize the domain growth of the CDW domains, we define a local CDW order parameter
    \begin{align}
        \phi_i\equiv\biggl(n_i-\frac{1}{4}\sum_j{}' n_j\biggr)\exp(i\mathbf{Q}\cdot\mathbf{r}_i),
    \end{align}
    where $n_i\equiv n_{i,\uparrow}+n_{i,\downarrow}$ denotes the total number electrons on site $i$, the prime in $\sum_j{}'$ indicates the summation is restricted to four nearest neighbors of site $i$, and $\mathbf{Q}=(\pi,\pi)$ corresponds to the ordering wave vector of the CDW. The intensity of the local order parameter measures the average local electron density difference with its nearest neighbors, and the sign distinguishes between the two types of checkerboard patterns of the CDW domain related by $\mathbb{Z}_2$ symmetry. Therefore, a uniform domain with a non-zero value of $\phi$ corresponds to a CDW domain characterized by the wave vector $\mathbf{Q}$.
    
    The snapshots of the simulation visualized by the local CDW order parameter $\phi_i$ are presented in Fig.~\ref{fig:3}. At the early stage (Fig.\ref{fig:3} {\bf a} and {\bf d}), we see the appearance of many small CDW domains across the system, indicting quick establishment of local equilibrium that saturates the CDW order parameter locally. As different small CDW domains start merging into large CDW domains (Fig.\ref{fig:3} {\bf b} and {\bf e}) , we can see the system with larger Hubbard $U$ seems to develop larger CDW domains compared to the system with smaller Hubbard $U$. This indicates that CDW domains in the system with larger Hubbard $U$ also grow faster. This trend becomes even more evident toward the late stage of the coarsening process (Fig.~\ref{fig:3} {\bf c} and {\bf f}). Although both systems have developed a large connected CDW domain, the system with smaller Hubbard $U$ has more disconnected CDW domains scattered around, and the sizes of these domains remain almost the same as they are at $t=200$, indicating the coarsening dynamics apparently slows down in the system with smaller Hubbard~$U$.
    
    To further characterize the growth of CDW domains under different Hubbard $U$ values, we compute the characteristic domain length $L(t)$ of the CDW domain during the coarsening process. This is done through computing the normalized equal-time correlation function
    \begin{align}
        C(r,t)=\frac{\langle \phi_i\phi_j\rangle-\langle\phi_i\rangle^2}{\langle \phi_i^2\rangle-\langle\phi_i\rangle^2}
    \end{align}
    with $r=|\mathbf{r}_i-\mathbf{r}_j|$. The correlation function is averaged over $70$ independent runs for each Hubbard $U$ value. Then we can compute the characteristic correlation length by $L(t)=\sum_r r|C(r,t)|^2/\sum_r |C(r,t)|^2$. The results for the time evolution of the characteristic domain length $L(t)$ under various Hubbard $U$ values are shown in Fig.~\ref{fig:4} {\bf a}. At $U/_{\mathrm{nn}}=0.3$, $L(t)$ grows very slowly and stops at the late stage, similar to the slow coarsening dynamics observed in the strong coupling limit of the Holstein model \cite{Cheng2023}. As the Hubbard $U$ increases, the growth of the characteristic domain length becomes faster and gradually approaches the Allen-Cahn $1/2$ power law, in agreement with the coarsening dynamics of the CDW in the weak coupling limit of the Holstein model \cite{Yang2025}. This implies that the Hubbard interaction effectively screens the Holstein coupling during the coarsening process, hence accelerating the CDW domain growth.

Furthermore, by rescaling $r/L(t)$, we can plot the correlation function at various time. If the coarsening dynamics preserves the dynamic scaling invariance, we should obtain a universal scaling function
    \begin{align}
        C(r,t) = f\left(\frac{r}{L(t)} \right),
    \end{align}
    independent of time \cite{Bray1994,Puri2009}. This is indeed the case at the late-stage of the coarsening process, as shown in Fig.~\ref{fig:4} {\bf b}-{\bf e}. After rescaling, the correlation functions collapse onto a single curve. This confirms that the late-stage coarsening dynamics of CDW domains from the HH model obeys dynamic scaling invariance.
    
\section{Discussion}
    The screening effect of the Hubbard interaction on the Holstein interaction in the coarsening dynamics of CDW domains raises important questions about the underlying mechanisms. Previous studies \cite{Clay2005,Kumar2008,Costa2020} have shown that an intermediate metallic state can emerge in the Hubbard-Holstein model when the Hubbard interaction is comparable to the Holstein interaction with the system being away from the adiabatic limit. While it might be tempting to attribute the observed screening effect to increased electron mobility due to the presence of the Hubbard interaction, it is crucial to note that in the adiabatic limit, the intermediate metallic phase vanishes. This suggests that other mechanisms are responsible for the screening effect of the Hubbard interaction.

    The screening effect of the Hubbard interaction has also been studied in the Hubbard model with random on-site potential $\epsilon_i$ \cite{Tanaskovic2003}. Under the framework of DMFT, it was found that the Hubbard interaction can effectively screen the on-site potential even in the DC limit by renormalizing the on-site disorder potential through the self-energy $\Sigma(\omega)$:
    \begin{align}
        \epsilon_i'=\epsilon_i +\Sigma_i(\omega=0).
    \end{align} 
    The strength of the screening can be derived by comparing the variance of the renormalized random potential, $\overline{\epsilon_i'^2}$, to the variance of the original potential, $\overline{\epsilon_i^2}$:
    \begin{align}
        \frac{\overline{\epsilon_i'^2}}{\overline{\epsilon_i^2}}=\frac{1}{(1+U\chi_{ii})^2},
    \end{align}
    where $\chi_{ii}=-\partial \langle n_i\rangle/\partial \epsilon_i$ denotes the local compressibility. Therefore, with finite Hubbard $U$, the renormalized random potential is always fractionally smaller than the original random potential on average.
    
   This explains the screening of the Holstein interaction $v_i=-gQ_i$, as an effective random on-site potential in the coarsening dynamics. The Holstein interaction is screened by the Hubbard interaction through the renormalization of the self-energy. Such renormalization process can also be seen from GA, where the renormalized Hamiltonian (\ref{eqn:qp}) gains an extra effective on-site potential that reduces the original Holstein interaction. Furthermore, the intrinsic competing nature of the Hubbard and Holstein interactions can be understood from a simple physics picture. The Hubbard interaction penalizes the double occupancy of electrons on a single site and favors a uniform charge distribution, while the Holstein interaction favors a nonuniform charge distribution to create a CDW state. As a result of this competition, the Hubbard interaction effectively screens the Holstein interaction, causing the coarsening dynamics of CDW domains to transition from the strong to the weak Holstein coupling limit in the adiabatic Hubbard-Holstein model.  

    In summary, we present a large-scale study of the non-equilibrium dynamics of CDWs in the adiabatic Hubbard-Holstein model, enabled by a machine learning force-field approach. The ML force field effectively utilizes the principle of locality and accurately predicts forces based on the local environment, providing a new approach for molecular dynamics simulations of large systems. Through our extensive large-scale simulations, we uncover an intriguing enhanced coarsening of CDW domains induced by electron correlation. To the best of our knowledge, our work is also the first systematic study of electron correlation effects on phase ordering dynamics. Our findings reveal the screening effect of the Hubbard interaction on the Holstein interaction in the coarsening dynamics of CDW domains, originating from the competing nature of these interactions even in the adiabatic limit. We show that the enhanced domain growth can be traced to a mechanism akin to the disorder screening due to electron correlation. These insights contribute to a better understanding of the fundamental mechanisms governing the interplay between electron-electron and electron-phonon interactions in strongly correlated systems and their impact on the non-equilibrium dynamics of CDWs.
    
\section{Methods}\label{Sec:Method}
    
\textbf{Gutzwiller approximation} -- The exact forces acting on the lattice degrees of freedom by electrons are computed from the Gutzwiller approximation. This is done through the construction of the Gutzwiller wave function \cite{Gutzwiller1963,Gutzwiller1964,Gutzwiller1965}
    \begin{align}
        |\Psi_G\rangle=\hat{\mathcal{P}}|\Psi_0\rangle=\prod_i\hat{\mathcal{P}_i}|\Psi_0\rangle,
    \end{align}
    where $\hat{\mathcal{P}}$ is the Gutzwiller operator that can be expressed as a product of on-site operators $\hat{\mathcal{P}}_i$ and $|\Psi_0\rangle$ is a Slater determinant of quasiparticle eigenstates which need to be optimized variationally. 

    With the help of slave-boson variables $\Phi_i$ \cite{Fabrizio2007,Lanata2008,Lanata2012}, we can map the expectation value
    $\langle \Psi_G|c_{i,\sigma}^\dagger c_{j,\sigma}|\Psi_G\rangle$ in the variational basis to the quasiparticle basis as $\mathcal{R}_{i,\sigma}\mathcal{R}_{j,\sigma}\langle \Psi_0|c_{i,\sigma}^\dagger c_{j,\sigma}|\Psi_0\rangle$ with
    \begin{align}
        \mathcal{R}_{i,\sigma}=\frac{\mathrm{Tr}\left(\Phi_i^\dagger \mathbb{M}_{i,\sigma}^\dagger\Phi_i \mathbb{M}_{i,\sigma}\right)}{\sqrt{n_{i\sigma}(1-n_{i\sigma})}},
    \end{align}
    where $\mathbb{M}_{i,\sigma}$ is the matrix representation of the electron annihilation operator $c_{i,\sigma}$ in the local basis, and $\Phi_i$ are slave-boson variables in the matrix form. Then we can write the variational energy of the electron part of the Hubbard-Holstein model in terms of $|\Psi_0\rangle$ and $\Phi_i$ as
    \begin{eqnarray}
        & &\mathcal{E}(|\Psi_0\rangle, \Phi_i)=\langle \Psi_G|\mathcal{H}_{\mathrm{elec}}|\Psi_G\rangle
        \\& & = \sum_{ij,\sigma}(t_
        {ij}-gQ_i\delta_{ij})\mathcal{R}_{i,\sigma}\mathcal{R}_{j,\sigma}\rho_{ij,\sigma}+U\sum_{i}\mathrm{Tr}(\Phi_i^\dagger\mathbb{D}_i\Phi_i),  \nonumber
    \end{eqnarray}
    where $t_{ij}=-t_{\mathrm{nn}}$ when $i$ and $j$ are nearest neighbors to each other, $\mathbb{D}_i$ is the matrix representation of the local double occupancy operator $\mathbb{D}_i\equiv n_{i,\uparrow}n_{i,\downarrow}$, and $\rho_{ij,\sigma}\equiv \langle\Psi_0|c_{j,\sigma}^\dagger c_{i,\sigma}|\Psi_0\rangle$ defines the single-particle density operator matrix. Moreover, the variational wave function is subject to the constraints $\langle \Psi_0|\mathcal{P}_i^\dagger\mathcal{P}_i|\Psi_0\rangle=\langle \Psi_0|\Psi_0\rangle=1$ and $\langle \Psi_0|\mathcal{P}_i^\dagger\mathcal{P}_ic_{i,\sigma}^\dagger c_{i,\sigma'}|\Psi_0\rangle=\langle\Psi_0|c_{i,\sigma}^\dagger c_{i,\sigma'}|\Psi_0\rangle$. 
    Then the minimization of the variational energy $\partial \mathcal{E}/\partial |\Psi_0\rangle=0$ and $\partial \mathcal{E}/\partial \Phi_i=0$ using Lagrange multipliers $\mu_i$ with the constraints gives two eigenvalues problems for quasiparticles and slave-bosons
    \begin{align}\label{eqn:qp}
        \mathcal{H}^{\mathrm{qp}}=\sum_{ ij,\sigma}(t_
        {ij}-gQ_i\delta_{ij})\mathcal{R}_{i,\sigma}\mathcal{R}_{j,\sigma}c_{i,\sigma}^\dagger c_{j,\sigma}+\sum_{i}\mu_i n_{i},
    \end{align}
    and
    \begin{align}
        \mathcal{H}_i^{\mathrm{sb}}=\sum_{\sigma}\frac{\Delta_{i,\sigma}\mathbb{M}_\sigma+\Delta^*_{i\sigma}\mathbb{M}^\dagger_\sigma}{\sqrt{n_{i,\sigma}(1-n_{i,\sigma})}}+\sum_{\sigma}\mu_{i,\sigma}\mathbb{N}_{i,\sigma}+U\mathbb{D}_i,
    \end{align}
    respectively. Here we define $\Delta_{i,\sigma}=\sum_{j}(t_{ij}-gQ_i\delta_{ij})\rho_{ij,\sigma}$ and $\mathbb{N}_{i,\sigma}$ is the electron number operator in the local basis. We solve these two Hamiltonians iteratively until the convergence reached for $|\Psi_0\rangle$ and $\Phi_i$.

\medskip

{\bf Symmetry-invariant descriptors} -- The construction of the descriptors follows from Ref.~\cite{Cheng2023}, where we utilize the irreducible presentation of the $D_4$ point group. On the square lattice, there are three types of local environment according to the irreducible representations of $D_4$. We can convert the lattice configuration $\{Q_0,Q_1,Q_2,\dots,Q_n\}$ on each local environment with the reference site $i$ into feature variables labeled by the corresponding irreducible representation $\{\bm{f}^\Gamma,\dots\}$ with the label $\Gamma$ denoting their irreducible representations. An example of one-dimensional irreducible representation is given by $f^{B_1}=a-b+c-d$, and two-dimensional irreducible representation is $\bm{f}^E=(a+b-c-d,a-b+c-d)$. These feature variables are transformed according to the irreducible representations of $D_4$, so the power spectrum of them, defined by $p^\Gamma\equiv|\bm{f}^\Gamma|^2$, is invariant under the symmetry transformation. However, we cannot only use the power spectrum of the feature variables as our symmetry-invariant descriptors because information provided by power spectrum is incomplete when the irreducible representation has the dimension more than one. Such information can be recovered from a more general set of invariants of the symmetry group called ``bispectrum coefficients" \cite{kondor2007}, which are triple products of the three irreducible representations. However, the bispectrum coefficients are a overcomplete set of variables describing the local environment, and many of them are redundant. Instead, we introduce the inner product of feature variables with the reference basis $\bm{f}^\Gamma_\mathrm{ref}$ for each distinct irreducible representation: $\eta^\Gamma\equiv\bm{f}^\Gamma\cdot\bm{f}^{\Gamma}_{\mathrm{ref}}/(|\mathbf{f}^\Gamma||\bm{f}^{\Gamma}_{\mathrm{ref}}|)$ \cite{Zhang2021}. The reference basis $\bm{f}_{\mathrm{ref}}^\Gamma$ is constructed from weighted average within the local environment. Moreover, we also need the relative phases between different irreducible representation variables. This is obtained from the inner product of the symmetry transformed reference basis $\mathcal{T}\bm{f}^\Gamma_{\mathrm{ref}}$ and a unit vector that is transformed within the same irreducible representation. For example, we can obtain $\eta^E_{\mathrm{ref}}=\mathcal{T}\bm{f}_{\mathrm{ref}}^E\cdot \mathbf{e}_1/|\bm{f}_{\mathrm{ref}}^E|$ with the unit vector $\mathbf{e}_1=(1,0)$ for the two-dimensional irreducible representation $E$.
Therefore, we obtain a set of generalized coordinates serving as symmetry-invariant descriptors $\left\{G_l\right\}=\{p^\Gamma,\eta^\Gamma,\eta_\mathrm{ref}^\Gamma\}$.

\medskip

{\bf Neural network and training} -- The ML force-field model is constructed and trained on PyTorch \cite{Paszke2019} with a six-layer neural network consisted of $45\times512\times256\times 64\times 16\times 1$ neurons. The number of neurons in the input layer is determined by the number of input number of descriptors $\{G_l\}$. The output layer is a single neuron which gives the predicted force. The loss function of the training process is defined as the mean square error of the local electronic forces
    \begin{align}
        \mathcal{L}=\frac{1}{N}\sum_{i}^N\left(F^{\mathrm{ML}}_i-F^{\mathrm{GA}}_i\right)^2.
    \end{align}
The ML model is trained separately under different Hubbard $U$ values. $300$ snapshots of forces and lattice configurations are contained in the training data set. In the training, the training data batch size is set to $1$ and $800$ epoches are applied to train the model with an adaptive learning rate $0.001$ in Adam optimizer \cite{Kingma2017}. 
    
\bigskip

\begin{acknowledgments}
This work was supported by the US Department of Energy Basic Energy Sciences under Contract No. DE-SC0020330. The authors acknowledge Research Computing at The University of Virginia for providing computational resources and technical support that have contributed to the results reported within this publication. 
\end{acknowledgments}
\bibliography{ML-HH.bbl}

\end{document}